\documentclass[amsmath,aip,jcp,preprint]{revtex4-1}

\usepackage{amsmath}
\usepackage{graphicx}
\usepackage{dcolumn}
\usepackage{bm}
\usepackage{slashbox}
\usepackage{multirow}
\usepackage{colortbl}

\begin{document}

\title{Ground state of small mixed helium  and spin-polarized tritium clusters: a quantum Monte Carlo study}

\author{P. Stipanovi{\'c}}
\affiliation{Faculty of Science, University of Split, HR-21000 Split,
Croatia}
\author{L. Vranje\v{s} Marki{\'c}}
\email{leandra@pmfst.hr}
\affiliation{Faculty of Science, University of Split, HR-21000 Split,
Croatia}
\affiliation{ Institut f\"ur Theoretische Physik, 
Johannes Kepler Universit\"at, A-4040 Linz, Austria}
\author{J. Boronat}
\affiliation{Departament de F\'\i sica i Enginyeria Nuclear, Campus Nord
B4-B5, Universitat Polit\`ecnica de Catalunya, E-08034 Barcelona, Spain}
\author{B. Ke\v{z}i{\'c}}
\affiliation{Faculty of Science, University of Split, HR-21000 Split,
Croatia}
\date{\today}

\begin{abstract}
We report results for the ground-state energy and structural properties of
small $^4$He-T$\downarrow$ clusters consisting of up to 4 T$\downarrow$ and
8 $^4$He atoms. These results have been obtained using very well-known
$^4$He-$^4$He and T$\downarrow$-T$\downarrow$ interaction potentials and
several models for the $^4$He-T$\downarrow$ interatomic potential. All the
calculations have been performed with variational and diffusion Monte Carlo
methods. It takes at least three atoms to form a mixed bound state. In
particular, for small clusters the binding energies are significantly
affected by the precise form of the $^4$He-T$\downarrow$ interatomic
potential but the stability limits remain unchanged. The only exception is
the $^4$He$_2$T$\downarrow$ trimer whose stability in the case of the
weakest $^4$He-T$\downarrow$ interaction potential is uncertain, while it seems stable for other potentials.
The mixed trimer $^4$He(T$\downarrow$)$_2$, a candidate for  Borromean state, is not bound.
All other studied clusters are stable. Some of the weakest bound clusters
can be classified as quantum halos, as a consequence of having high
probability of being in a classically forbidden region.    
\end{abstract}


\maketitle

\section{Introduction}\
There are many similarities between spin-polarized tritium (T$\downarrow$)
and $^4$He atoms. Both are bosons, have a small mass and a weakly attractive
interaction potential at large distances.  Just like $^4$He, bulk
T$\downarrow$ remains liquid in the limit of zero temperature and zero
pressure. In 1976, Stwaley and Nosanow suggested that T$\downarrow$ should
behave very much like liquid $^4$He and therefore constitute a second
example of bosonic superfluid.~\cite{StwaleyNosanow} Many similarities
between the two systems have been confirmed by recent quantum Monte
Carlo~\cite{tritium} results.  For instance, bulk T$\downarrow$ is a
superfluid at zero temperature, with a condensate fraction $n_0$=0.129(3) at
the equilibrium density. Miller examined
the bulk $^4$He-T$\downarrow$ system as an example of binary mixtures using a variational ansatz and
found  that it prefers to be completely phase-separated.~\cite{miller}

$^4$He clusters have been extensively studied, both theoretically and
experimentally, exhibiting nanoscopic superfluidity. Due to the
similarities between $^4$He and T$\downarrow$, a similar behavior of
spin-polarized tritium clusters is expected. So far, T$\downarrow$ clusters
have been investigated only theoretically. Small clusters of T$\downarrow$
were for the first time studied by Blume \textit{et al.}~\cite{blume1}. The
smallest bound T$\downarrow$ cluster is the trimer, showing characteristics
of a Borromean or halo state.~\cite{Fedorov,Jensen} Its tiny binding energy
of only around 4 mK has been later confirmed by other authors
\cite{salci,smallhyd}.  The extension of the study to larger
clusters~\cite{Tlargeclusters} confirmed that T$\downarrow$ clusters are
much more weakly bound and diffuse than the $^4$He clusters with the same
number of atoms. The stability of mixed clusters of
T$\downarrow$~\cite{smallhyd,tritiumHD} with spin-polarized hydrogen
(H$\downarrow$) and spin-polarized deuterium (D$\downarrow$), in particular
small ones, has also been investigated. For the clusters with D$\downarrow$
the stability limits depend on the number of D$\downarrow$ atoms and the
occupation of its nuclear spin states. On the other hand, due to the small
mass of H$\downarrow$, it has been shown that even 60 T$\downarrow$ atoms are not
enough to bind one H$\downarrow$ atom.

Mixed $^4$He-T$\downarrow$ clusters have not been studied yet.
T$\downarrow$ has almost the same mass as $^3$He, but it is a boson, so it
will exhibit different physical properties. Because of the small mass of
both $^4$He and T$\downarrow$ and  weak attractive parts of the interaction
potentials it is expected that the smallest clusters   could be extremely
weakly bound. If $^4$He(T$\downarrow$)$_2$ is bound, it would be an example
of yet another Borromean state, because none of its subsystems are bound.
The only mixed molecular system for which Borromean state has been
predicted theoretically so far is $^3$He$_2$K.~\cite{LiGo} Recently, small
mixed helium-hydrogen clusters  $^4$He$_2$H, $^4$He$_2$H$^-$, $^4$HeH$_2$
have been studied as
well.~\cite{LiLin,Casalegno,Gianturcoetal01,LiuRoy,HeH2}  

Weakly bound clusters are also candidates for Efimov states,~\cite{Efimov}
which are predicted to exist also in mixed systems. Interest in these
states and other universal binding properties of small clusters has been
significantly intensified in the last couple of years  by the research in
the field of cold gases. After the experimental detection of an Efimov
state in  an  ultracold gas of cesium atoms,~\cite{cesium} giant Efimov
trimers have been detected in other
systems~\cite{Ottenstein08,Huckans09,Gross09} including those of mixed
species.~\cite{Barontini09}  Later work was devoted
to the study of the Efimov spectrum~\cite{Zaccanti09}  as well as to the prediction~\cite{vonStecher09}
 and detection~\cite{Ferlaino10} of giant tetramers. Furthermore, a recent study
of weakly bound bosonic clusters has found a  series of universal cluster
states, that can be qualitatively interpreted by adding one particle at a
time to an Efimov trimer.~\cite{vonStecher10}

Although the physics of molecular weakly bound clusters is not the same as
the one of ultracold gases, it is interesting to investigate if similar
behavior of  few-body states appear, indicating the universality. Universal
physics is also connected to physics of quantum
halos,~\cite{RMPhalos,HannaBlume,Brunnian} defined as bound states of
clusters of particles with a radius extending well into the classically
forbidden region. It is therefore worth studying which small molecular
clusters qualify as quantum halos.

In this work, we report the ground-state energy of clusters having up to 8 $^4$He and 4 T$\downarrow$ atoms, as well as their structural properties, obtained with the diffusion Monte Carlo (DMC) method. 
In Sec. II, we report briefly the DMC method and discuss the trial wave
functions used for importance sampling.  Sec. III reports the results obtained by the DMC simulations. Finally, Sec. IV comprises a summary of the
work and an account of the main conclusions.

\section{Method}
\label{method}

\subsection {Interaction potentials}

 We have
modeled the $^4$He-$^4$He interactions with  the Aziz HFD-B(He) interaction
potential~\cite{HFD}. Some calculations have also been performed with the
Korona \textit{et al.} SAPT potential.~\cite{Korona} 

The interatomic interaction between tritium atoms is described with the
spin-independent central triplet pair potential $b$$^3\Sigma_u^+$, which was  
determined in an essentially  exact way by Kolos and Wolniewicz.~\cite{kolos} 
As in our recent DMC calculations of bulk H$\downarrow$ and T$\downarrow$,\cite{hydrogen,tritium} we have  used the recent extension of Kolos and Wolniewicz data to larger interparticle distances by Jamieson \textit{et al.}
(JDW).~\cite{jamieson} The potential is finally constructed using a cubic spline interpolation of JDW data, which is smoothly connected to the long-range behavior of the T$\downarrow$-T$\downarrow$  potential as
calculated by Yan \textit{et al.}~\cite{yan}
The JDW potential used in the present work has a core diameter
$\sigma=3.67$~\AA\, and a minimum of $-6.49$ K at a distance $4.14$ \AA.  We have previously verified that the addition of mass-dependent adiabatic corrections (as calculated by Kolos and Rychlewski~\cite{kolos2}) to the JDW potential does not change the energy of bulk spin-polarized tritium.~\cite{tritium} It is worth mentioning that within the Born-Oppenheimer approximation it has been explicitly shown that in the spin-aligned electronic state, tritium 
nuclei behave as effective bosons.~\cite{freed}

\begin{figure}
\centering
        \includegraphics[width=8.5 cm,angle=0]{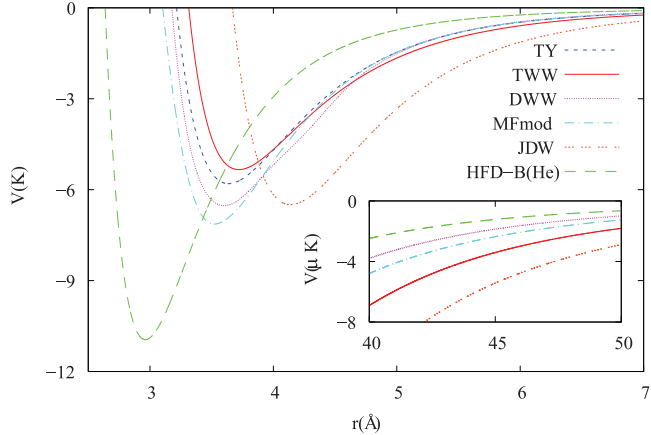}
                \caption{Comparison among interaction potentials  between $^4$He-$^4$He (HFD-B(He)~\cite{HFD}), $^4$He-T$\downarrow$ (TY~\cite{TY}, TWW~\cite{TWW}, DWW~\cite{Das}, MFmod~\cite{MF,MFmod}) and T$\downarrow$-T$\downarrow$ (JDW~\cite{jamieson}). For long distances, TY and MFmod have basically the same behavior. }
                \label{figpot}
\end{figure}

Several forms of $^4$He-T$\downarrow$ interaction potentials are available
in the literature, including those resulting from {\it ab initio}
calculations and semiempirical potentials. In 1984, Jochemsen  {\it et
al.}~\cite{Jochemsen}  proposed a semiempirical potential (R2) after
discussing available potentials until that time and comparing theoretical
data with diffusion experiments at low temperature. That potential is
essentially the same as an older potential by Das {\it et al.}~\cite{Das}
(DWW), which the  authors of Ref. \onlinecite{Jochemsen} themselves
preferred to use. The most sophisticated {\it ab initio} calculation is
that of Meyer and Frommhold from 1994 (MF), who obtained good agreement
with experiment, except at very low temperatures.~\cite{MF} Subsequently,
Chung and Dalgarno~\cite{MFmod} introduced slight modifications in the
short-range repulsive part of the MF potential that resulted in a better
agreement with diffusion measurements at low temperatures (MFmod). In both
MF and MFmod potentials,  the construction
of the potential-energy curve at large separations and the dispersion
component that determines the long-range behavior in MF and MFmod is of
high precision.~\cite{yan} All of our calculations have been performed with
both the DWW and the MFmod potential. In addition, for some selected
clusters we have performed calculations using some other forms of available
potentials. Specifically, Toennies {\it et al.}~\cite{TWW} have proposed
the Lennard-Jones 6-12 potential (TWW) whose parameters have been
determined using data from low energy elastic scattering. TWW has the
largest core, smallest depth and the strongest long-range part of all the
studied potentials. Due to its simplicity it has been used also in some
very recent calculations, e.g. by Krotscheck and Zillich;~\cite{KroZill}
these authors have found that, in the bulk liquid, the binding energies
obtained with TWW differ less than 5\% from those that are obtained with
the DWW potential. Finally, we include the MF potential and the potential
by Tang and Yang~\cite{TY} (TY), who used the Tang-Toennies model. The
comparison of these different potentials is presented in Figure
\ref{figpot}. The strongest MFmod potential has a core diameter of
$\sigma=3.10$~\AA\, and a minimum of $-7.14$ K at a distance $3.52$ \AA.
The MF potential would  be the same in the
 scale of the  figure because the only
differences  are  in the repulsive part. The DWW interaction has a core
diameter of $\sigma=3.18$~\AA\, and a minimum of $-6.53$ K at a distance
$3.60$ \AA, while the weakest TWW potential has a core diameter of
$\sigma=3.31$~\AA\, and a minimum of $-5.34$ K at a distance $3.72$ \AA. In
Fig. \ref{figpot}  the  $^4$He-$^4$He and T$\downarrow$-T$\downarrow$
interaction potentials have been included for comparison as well. Of all
the interaction potentials,  the  T$\downarrow$-T$\downarrow$
 one  has the largest
core and the most attractive long-range part. 

\subsection {Diffusion Monte Carlo method}

The ground-state properties of the clusters have been  studied  using the DMC method, whose starting point is the many-body Schr\"odinger equation written in imaginary time,
\begin{equation}
-\hbar \frac{\partial \Psi(\bm{R},t)}{\partial t} = (H- E_{\text r}) \Psi(\bm{R},t) \ ,
\label{srodin}
\end{equation}
where $E_{\text r}$ is a constant acting as a reference energy and
$\bm{R} \equiv (\bm{r}_1,\ldots,\bm{r}_N)$ collectively denotes particle positions.

The $N$-particle Hamiltonian $H$ for the cluster $^4$He$_n$(T$\downarrow$$)_m$, $n+m=N$ is
\begin{equation}
 H = - \sum_{i=1}^{N} \frac{\hbar^2}{2m_i}\bm{\nabla}_i^2 + \mathop{\sum_{i,j=1}^{n}}_{i<j} V_{HeHe}(r_{ij})+\sum_{i=1}^{n}\sum_{j=n+1}^{n+m} V_{HeT}(r_{ij}) + \mathop{\sum_{i,j=n+1}^{n+m}}_{i<j} V_{TT}(r_{ij}) \ ,
\label{hamilto}
\end{equation}
where $V_{HeHe}(r)$, $V_{HeT}(r)$ and $V_{TT}(r)$ are the interaction potentials between the different components of the mixture.   

In DMC,  the Schr\"odinger
equation (\ref{srodin}) is solved stochastically  by multiplying $\Psi(\bm{R},t)$ with  $\psi(\bm{R})$, a trial wave function used for importance sampling, and rewriting Eq. (\ref{srodin}) in terms of the mixed distribution $\Phi(\bm{R},t)=
\Psi(\bm{R},t) \psi(\bm{R})$. Within the Monte Carlo framework, $\Phi(\bm{R},t)$ is represented by a set of \textit{walkers} $\bm{R_i}$. 
In the limit $t \rightarrow \infty$, only the lowest energy eigenfunction, not orthogonal to $\psi(\bm{R})$, survives and then the sampling of the ground state is effectively achieved. Apart from statistical uncertainties, the energy of a $N$-body bosonic system is exactly calculated.

In the present simulations, the trial wave function has been written as a product of Jastrow two-body correlation functions between all the pairs,
\begin{equation}
\psi_{\text J}(\bm{R}) = \mathop{\prod_{i,j=1}^{n}}_{i<j} f_{HeHe}(r_{ij})\mathop{\prod_{i,j=n+1}^{n+m}}_{i<j} f_{TT}(r_{ij})\prod_{i=1}^{n}\prod_{j=n+1}^{n+m}f_{HeT}(r_{ij}),
\label{trial}
\end{equation}
where $f_{HeHe}$ describes two-body correlations between helium atoms, $f_{TT}$ describes two-body correlations between spin-polarized tritium atoms and $f_{HeT}$ accounts for the $^4$He-T$\downarrow$ pairs.
Different types of two-body correlation function $f(r)$ have been tested, 
\begin{equation}
f_1(r)=\exp\left[-\left(\frac{b}{r}\right)^{5}-sr\right] \ ,
\label{eq:trial1}
\end{equation}
\begin{equation}
f_2(r)=\exp\left[-\left(\frac{\alpha}{r}\right)^{\gamma}-s_1 r\right]/r \ ,
\label{eq:trial2}
\end{equation}
\begin{equation}
f_3(r)=\exp [-b_1 \exp(-b_2r )-s_2 r] 
\label{eq:trial3}
\end{equation}
where $b$, $s$, $\alpha$, $\gamma$, $s_1$, $b_1$, $b_2$ and $s_2$  are variational parameters. 
The optimization of the trial wave functions has been done for all clusters by means of the  variational Monte Carlo (VMC) method.
For clusters having less than 5 atoms it has been very difficult to obtain good quality VMC results so different forms have been tried out to ascertain that DMC gives the same value of the energy for different guiding wave functions. 

For $^4$He-$^4$He  correlations, 
 a  set of calculations for all clusters has
been done using   $f_1(r)$ (\ref{eq:trial1}), with the parameter $b$=2.6
\AA\, and $s$ from 0.08 to 0.13 \AA$^{-1}$. For some smaller clusters,
$f_2(r)$ (\ref{eq:trial2}) has also been used, with optimal parameters  in
the range $\alpha\in\left[2.75,2.82 \right]$ \AA, $\gamma\in\left[4.1,4.7
\right]$, $s_1\in\left[0.001,0.015 \right]$ \AA$^{-1}$. For
T$\downarrow$-T$\downarrow$ two-body correlations, the form $f_3(r)$
(\ref{eq:trial3}) has been chosen for all the clusters, based on the
previous experience with pure and mixed T$\downarrow$
clusters.~\cite{smallhyd} The optimal parameters have been obtained in the
range: $b_1$ from 82 to 94, $b_2$ from 1.23 to 1.24 \AA$^{-1}$, and $s_2$
from 0.04 to 0.06 \AA$^{-1}$, where smaller values of $s$ correspond to
larger clusters. The only exceptions are the  clusters
$^4$He$_2$(T$\downarrow$)$_2$ and $^4$He(T$\downarrow$)$_{3,4}$ for which
better energies were obtained with $b_1=200$, $b_2=1.5$ \AA$^{-1}$ and
$s_2=0.065$ \AA$^{-1}$. In addition, for some of the smallest clusters
 variational calculations have also been done with the form $f_1(r)$ (\ref{eq:trial1}), where optimal
parameters have been found around $\alpha=4.2$ \AA, $\gamma \in \left[3.15,
3.82\right]$ and $s_1\approx$ 0.001 \AA$^{-1}$.

The  optimization of the  $^4$He-T$\downarrow$ correlations  has   been the most demanding:  it  has been done for several potentials and forms
of trial wave functions. First, for the DWW potential, $f_1(r)$ has been
used with $b$ around 3 \AA\, and $s$ ranging from 0.04 to 0.06 \AA$^{-1}$.
With $f_1(r)$ and clusters with more than 5 atoms the VMC energies approach
the  80\% of the DMC values. Therefore, for the smallest clusters, 
$f_2(r)$ has been tried as well, with resulting parameters $\alpha \in
\left[3.59, 3.69\right]$ \AA, $\gamma \in \left[3.08,3.20\right]$,
$s_1=0.001$ \AA$^{-1}$. In conclusion, VMC energies reached more than 40\%
of the DMC ones in all cases except $^4$He$_2$T$\downarrow$, where negative
energy could not be obtained at the VMC level for the DWW potential. For
this cluster, the best set of parameters for the function $f_2(r)$ is
$\alpha=3.62$ \AA, $\gamma=3.1$, $s_1=0.001$ \AA$^{-1}$, although it
arrives to the same DMC value for other sets of parameters. In addition,
$^4$He-T$\downarrow$ correlations for that cluster have also been modeled
with $f_3(r)$ (\ref{eq:trial3})  ($b_1=300$, $b_2=1.93$ \AA$^{-1}$ and
$s_2=0.05$ \AA$^{-1}$) and despite significant differences at the VMC
level, DMC has arrived to the same value of the energy. Variational
parameters differ  only  slightly for other types of $^4$He-T$\downarrow$
interaction potentials, except
for  trimers and tetramers,  where  the  differences are somewhat larger.

We have verified that 1000 walkers are enough for excluding the bias coming from the size of the population ensemble used in a simulation. The same conclusion emerged also from previous experience in pure and mixed T$\downarrow$ clusters. The only exception is again the $^4$He$_2$T$\downarrow$ cluster, for which a larger number of walkers has been necessary: 2000  for the DWW potential and 10000 for the TWW potential.

In order to eliminate bias coming from the time-step value used in
simulations, all calculations have been performed with several time-steps
($\Delta t$) which assume values within the interval $3\times 10^{-4} -
5\times 10^{-3}$ K$^{-1}$, where minimum and maximum values in that
interval have been adjusted depending on the size of the cluster.  The 
ground-state energies for different time-steps have then been extrapolated
to $\Delta t \rightarrow 0$. In accordance with the DMC method used in this
work, which is accurate to second order in the time step,~\cite {boro} the
extrapolation is made with a quadratic function.
 
The potential energies and distribution functions have been obtained using
pure estimators.~\cite{pures}  The asymptotic block size for each cluster
and type of $^4$He-T$\downarrow$ interaction potential has been determined
from the behavior of the potential energy versus the block size. For most
of the clusters, 2500 steps  per block  have been enough to arrive to the asymptotic
value of the potential energy. In the case of smaller clusters, having less
than five atoms, the block size  grows  to 10000-20000 steps reflecting
the low quality of the trial wave functions. Even 50000 steps in the block
have been needed for  the  TWW potential. However, for blocks of this size the
same values of pure estimators for different types of trial wave functions
have been obtained and therefore DMC estimations of these magnitudes are
unbiased.  

\section{Results}

\begin{figure}
	\centering
		\includegraphics[width=8.5cm]{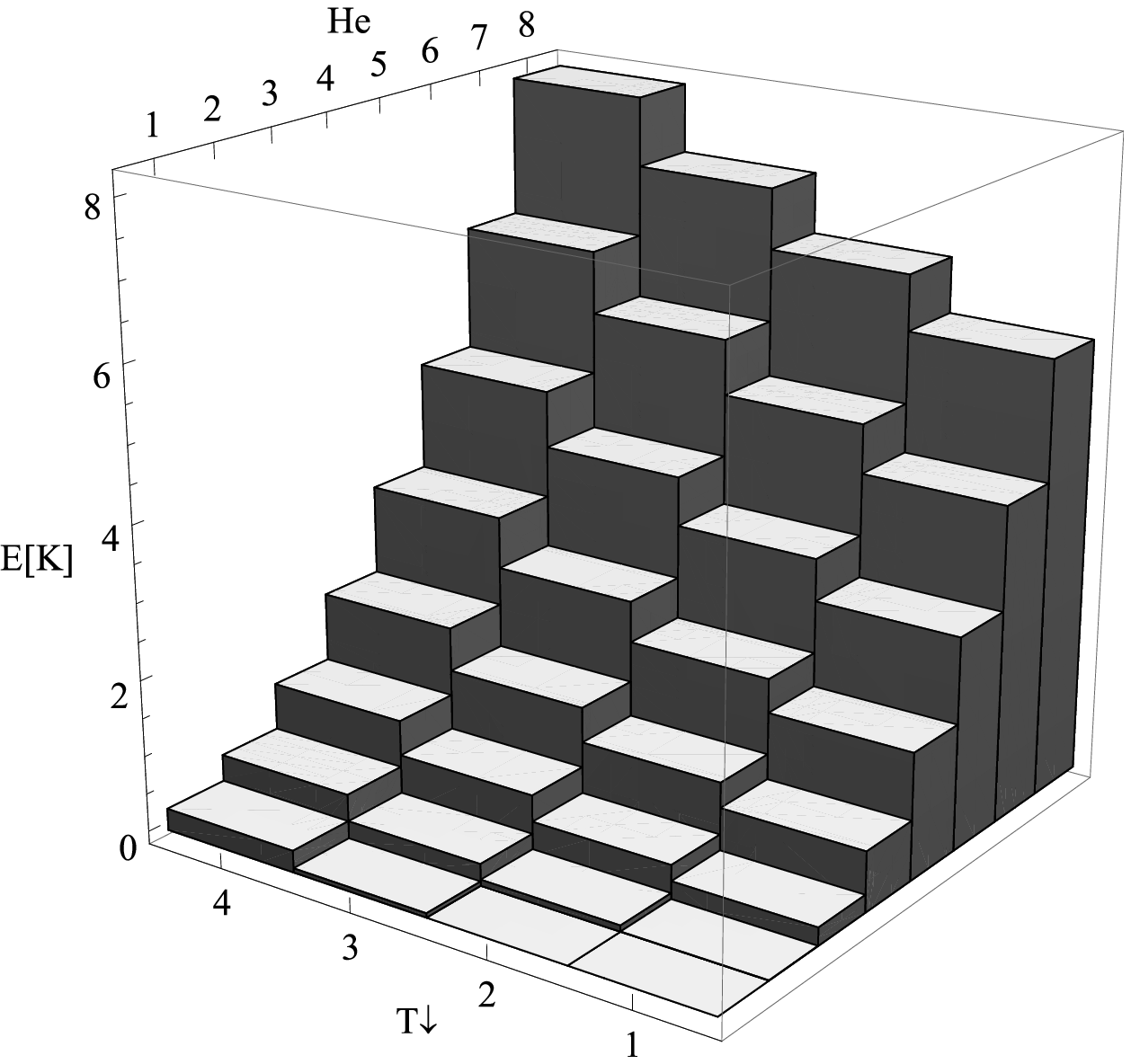}
	\caption{The absolute values of the ground-state energies are presented as 
	a function of the number of $^4$He and T$\downarrow$ atoms, when using the 
	DWW~\cite{Das} potential to model the $^4$He-T$\downarrow$ interaction.}
	\label{fig:endas}
\end{figure}

Ground-state energies of all the studied clusters for two types of $^4$He-T$\downarrow$ interaction potentials are presented in Table \ref{tab:energies1}. Energy in absolute value as a function of the number of $^4$He and T$\downarrow$ atoms is presented in Fig. \ref{fig:endas} for the DWW interaction potential. In addition, calculations which have been made for several selected clusters with other types of potentials are presented in Table \ref{tab:energies2}. 

\begin{table}[htbp]
  \centering
\newcolumntype{H}{>{\columncolor[gray]{0.8}[0.9\tabcolsep]}r}
  \caption{\label{tab:energies1} Ground-state energies (in mK) of the studied $^4$He-T$\downarrow$ clusters. In each cell top result and bottom result are obtained using the DWW~\cite{Das} and  MFmod~\cite{MFmod} $^4$He-T$\downarrow$ interactions, respectively.}
    \begin{tabular}{c|r|r|r|r}

    \backslashbox{$^4$He}{T$\downarrow$}   & \multicolumn{1}{c|}{1}     & \multicolumn{1}{c|}{2}    
     & \multicolumn{1}{c|}{3}     & \multicolumn{1}{c}{4} \\ \hline
    \multirow{2}{*}{1}  &       &       & -56(3) & -285(5) \\
             &       &       & \multicolumn{1}{H|}{-92(5)} & \multicolumn{1}{H}{-355(5)} \\ 
	     \hline
    \multirow{2}{*}{2} & -3.0(2) & -80(6) & -284(6) & -635(6) \\
                       & \multicolumn{1}{H|}{-16.6(8)} &  \multicolumn{1}{H|}{-135(9)} & 
		        \multicolumn{1}{H|}{-406(8)} &  \multicolumn{1}{H}{-819(9)} \\ \hline
    \multirow{2}{*}{3}  & -235(10) & -444(5) & -790(10) & -1248(10) \\
                        &  \multicolumn{1}{H|}{-290(16)} &  \multicolumn{1}{H|}{-577(5)} &  
			\multicolumn{1}{H|}{-1007(6)} &  \multicolumn{1}{H}{-1552(6)} \\ \hline
    \multirow{2}{*}{4} & -790(5) & -1120(6) & -1577(5) & -2145(11) \\
                       &  \multicolumn{1}{H|}{-883(4)} &  \multicolumn{1}{H|}{-1318(4)} &  
		       \multicolumn{1}{H|}{-1885(4)} &  \multicolumn{1}{H}{-2560(6)} \\ \hline
    \multirow{2}{*}{5} & -1664(5) & -2105(6) & -2640(6) & -3310(14) \\
                       &  \multicolumn{1}{H|}{-1784(5)} &  \multicolumn{1}{H|}{-2356(8)} &  
		       \multicolumn{1}{H|}{-3014(7)} &  \multicolumn{1}{H}{-3825(8)} \\ \hline
    \multirow{2}{*}{6} & -2785(7) & -3335(9) & -3975(6) & -4715(13) \\
                       &  \multicolumn{1}{H|}{-2925(8)} &  \multicolumn{1}{H|}{-3636(8)} & 
		        \multicolumn{1}{H|}{~-4433(10)} &  \multicolumn{1}{H}{-5320(8)} \\ \hline
    \multirow{2}{*}{7} & ~-4145(11) & -4785(7) & -5510(7) & ~-6330(13) \\
                       &  \multicolumn{1}{H|}{-4302(6)} &  \multicolumn{1}{H|}{-5121(8)} & 
		       \multicolumn{1}{H|}{ -6008(10)} &  \multicolumn{1}{H}{~-7015(8)} \\ \hline
    \multirow{2}{*}{8} & -5725(8) & -6455(11) & -7250(10) & -8155(6) \\
                       &  \multicolumn{1}{H|}{~-5908(8)} &  \multicolumn{1}{H|}{~-6834(12)} & 
		        \multicolumn{1}{H|}{~-7818(10)} &  \multicolumn{1}{H}{-8916(14)} \\ \hline
    \end{tabular}
\end{table}

As shown in Table \ref{tab:energies1}, it takes at least three atoms to
form a mixed bound state. The weakest bound cluster is
$^4$He$_2$T$\downarrow$. However, the value of its binding energy depends
strongly on the type of the interaction potential. For the DWW, MF and
MFmod potentials, it is lower than the energy of the $^4$He$_2$ dimer,
which is -1.66 mK, as calculated in Ref. \onlinecite{mol3d}. For the two weaker
potentials (TWW, TY) the obtained energy is within the errorbar equal to
the energy of the $^4$He$_2$ dimer. Thus, in the case of the latter two
potentials it is unclear  whether  this cluster is stable with respect to the
separation in the $^4$He$_2$ dimer and a free T$\downarrow$ atom. On the
other hand, for the potentials in which $^4$He$_2$T$\downarrow$  is surely
bound the value of the binding energy varies from -3.0(2) mK to -16.6(8)
mK. Such small energies  result from  a huge cancellation between kinetic
and potential energies. For instance, in case of the DWW interaction
potential the potential energy of the $^4$He$_2$T$\downarrow$ is
$E_p=-300(20)$ mK, as determined using pure estimators. The kinetic energy,
determined as a difference between the total and potential energy is then
297(20) mK. A similar system, $^4$He$_2$H has recently been investigated
using an interaction potential which is in between the DWW and
MFmod~\cite{Cvetko} and no bound states have been
found.~\cite{LiLin,LiuRoy,HeH2} This is very much in agreement with the
present result, since the three times lower mass of hydrogen increases the
kinetic energy of the system, which already for tritium almost cancels the
potential energy. In Ref. \onlinecite{LiLin} other isotopic combinations
have been studied as well using hyperspherical coordinates in the adiabatic
approximation. One bound state has been found for $^4$He$_2$T system, with
the energy of -6.8 mK. Since the authors do not use the same form of the
$^4$He-$^4$He interaction potential it is not possible to directly compare
this value with our results; however we obtain qualitative agreement.
\begin{table}
\begin{center}
\caption{\label{tab:energies2} Ground-state energies (in mK) of investigated $^4$He-T$\downarrow$ clusters.} 
\begin{tabular}{c|c|c|c|c} 
\hline
\hline
\multicolumn{2}{c|}{Cluster} &
\multicolumn{3}{c}{$^4$He-T$\downarrow$ interaction potential }\\ \hline
$^4$He & T$\downarrow$ & TWW & TY &   MF  \\ \hline
2  & 1  & -1.3(3)  & -1.4(4)  & -13.9(8)\\
3  & 2  & -254(12) & -288(15) & -522(4)  \\
3  & 4  & -781(11)  & -877(7)  & -1434(8)  \\ 
4	 & 2	& -834(6) &	-886(6) &			-1243(5)  \\
4	 & 4	& -1503(10) &	-1633(7)  &			-2402(10)  \\ 
6	&1	&-2579(6)&	-2605(7) 	& -2863(9)	 \\
6	&4	&~-3826(10) &~-3990(10) & ~-5075(11)  \\  \hline
\hline
\end{tabular} 
\end{center}
\end{table}   

An extensive search for a bound state of the other mixed trimer $^4$He(T$\downarrow$)$_2$, a possible Borromean state, has been performed. However, for all of the studied potentials and different types of trial wave functions, the energies in the DMC calculations remained positive. Moreover, in the course of the simulation, particles were going more and more away from each other. This case can be compared with the $^4$He$^3$He$_2$ cluster which has basically the same mass, but interacts with a stronger potential and is likewise not bound.~\cite{EsryLinGreene}

\begin{figure}
	\centering
		\includegraphics[width=8.5cm]{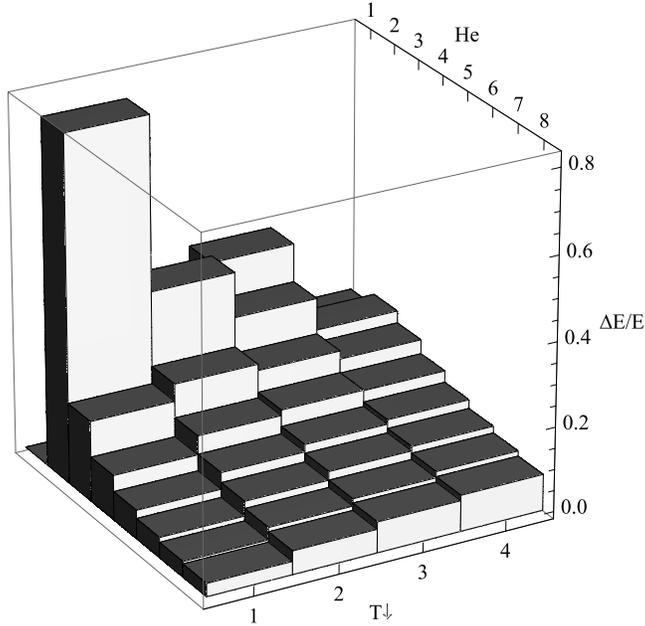}
	\caption{\label{fig:Efrac}The difference between ground-state energies in calculations with the MFmod~\cite{MFmod} and DWW~\cite{Das} potentials, divided by the ground-state energy with the MFmod potential, $\Delta E/E=(E_{MFmod}-E_{Das})/E_{MFmod}$, as a function of the number of $^4$He and T$\downarrow$ atoms in the cluster.}
\end{figure}

All the other studied mixed clusters with up to four T$\downarrow$ and
eight $^4$He atoms are bound. In absolute value, the energy grows with the
addition of both $^4$He and T$\downarrow$ atoms, but it takes about two
T$\downarrow$ atoms to achieve the same increase in binding as
 with the
addition of one $^4$He atom, as can be seen in Fig. \ref{fig:endas}.
Significant cancellation between kinetic and potential energies persists
for almost all clusters, with total energies being 1\% to 20\% of the
potential energy for the largest cluster. The differences in the binding
energies for different potentials
 are lowered with the increase
of the cluster size and in particular with combined lowering of the
fraction of T$\downarrow$ atoms in it. This behavior can be more easily
seen in Fig. \ref{fig:Efrac} which shows the difference between
ground-state energies in calculations with MFmod and DWW potentials,
divided by the ground-state energy with the MFmod potential
($E_{MFmod}-E_{Das})/E_{MFmod}$, as a function of the number of $^4$He and
T$\downarrow$ atoms in the cluster.   Qualitatively, the stability of the
clusters, except for $^4$He$_2$T$\downarrow$, is not sensitive to the forms
of the interaction potentials. However, the relationship of binding
strengths between different clusters is not always the same for different
potentials. For example, $^4$He$_7$(T$\downarrow$)$_2$ is more strongly
bound than $^4$He$_6$(T$\downarrow$)$_4$ for the DWW potential, while for
the MFmod one this relationship is reversed.  Some calculations have been
performed using another form of the $^4$He-$^4$He interaction
potential~\cite{Korona}. The resulting energies are basically  equal
 within the
errorbars. For example, for $^4$He$_2$T$\downarrow$ with the HFD-B(He)
potential and DWW one obtains -3.0(2) mK, and with SAPT~\cite{Korona} and
DWW -3.3(2) mK.  The  ground-state energy of  the   larger cluster
$^4$He$_8$T$\downarrow$ is -5770(10) mK with SAPT,  and -5725(8) mK with
HFD-B(He). 

The smaller clusters are especially interesting. For example, $^4$He(T$\downarrow$)$_{3,4}$ has no bound two-particle subsystem, while out of all six two-particle interactions in $^4$He$_2$(T$\downarrow$)$_2$  only the one between two $^4$He atoms would give a bound subsystem. The latter cluster is similar to $^4$He$_2$$^3$He$_2$,  which authors of Ref. \onlinecite{expmix} have called pseudo-Borromean state because of the fact that it has also only one bound subsystem.

\begin{figure}
	\centering
		\includegraphics[width=8.5cm]{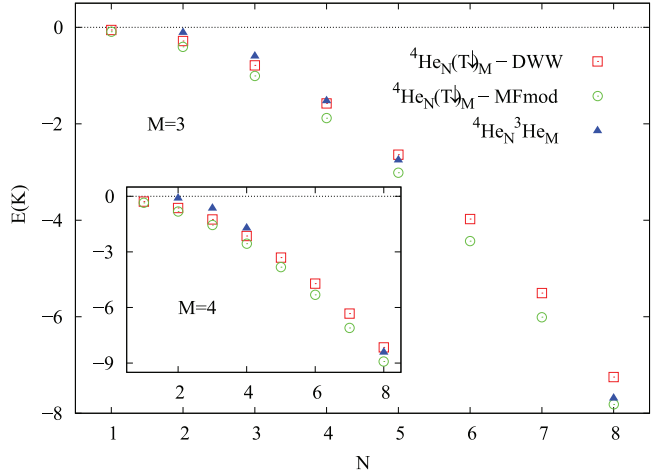}
	\caption{\label{fig:encomphe}Comparison of the ground-state energies of 
	$^4$He$_N$(T$\downarrow$)$_M$ clusters for two $^4$He-T$\downarrow$ 
	interaction potentials with the ground-state energies of $^4$He$_N$
	$^3$He$_M$ 
	clusters for different $N$. The larger figure shows results for $M=3$, while results for 
	$M=4$ are shown in the inset. Axis labels are the same on both figures. 
	Ground-state energies of $^4$He$_N$ $^3$He$_M$ clusters   are taken from Ref. \onlinecite{mol3d} and \onlinecite{GN2003}. }
\end{figure}

The comparison between $^4$He$_n$(T$\downarrow$)$_m$ and $^4$He$_n$$^3$He$_m$ is interesting because both types of clusters have essentially the same mass, but differ in the interaction potentials and the statistics. The interaction potential between two helium atoms is the same for all combinations of isotopes and is stronger than both the $^4$He-T$\downarrow$ and T$\downarrow$-T$\downarrow$ one. Thus, as expected, the clusters $^4$He$_n$$^3$He$_m$ are more strongly bound than $^4$He$_n$(T$\downarrow$)$_m$ for $m$=1,2. However, for more than two $^3$He atoms the fermion statistics comes into play. Thus for example $^4$He$^3$He$_{3,4}$ and $^4$He$_2$$^3$He$_{3,4}$ are unbound, but the clusters $^4$He(T$\downarrow$)$_{3,4}$ and $^4$He$_2$(T$\downarrow$)$_{3,4}$ have bound states. For larger clusters, despite the fermionic nature of $^3$He, $^4$He$_n$$^3$He$_m$ show tendency to be more strongly bound than $^4$He$_n$(T$\downarrow$)$_m$, for increasing $n$, as can be seen in Fig. \ref{fig:encomphe} for the DWW potential, but is not yet clear for the MFmod. The precise $m$ for which this transition occurs can not be predicted because the energies of $^4$He$_n$$^3$He$_m$ are not calculated for all $n$ and $m$ with the HFD-B(He) interaction potential.
\begin{figure}
	\centering
		\includegraphics[width=8.5cm]{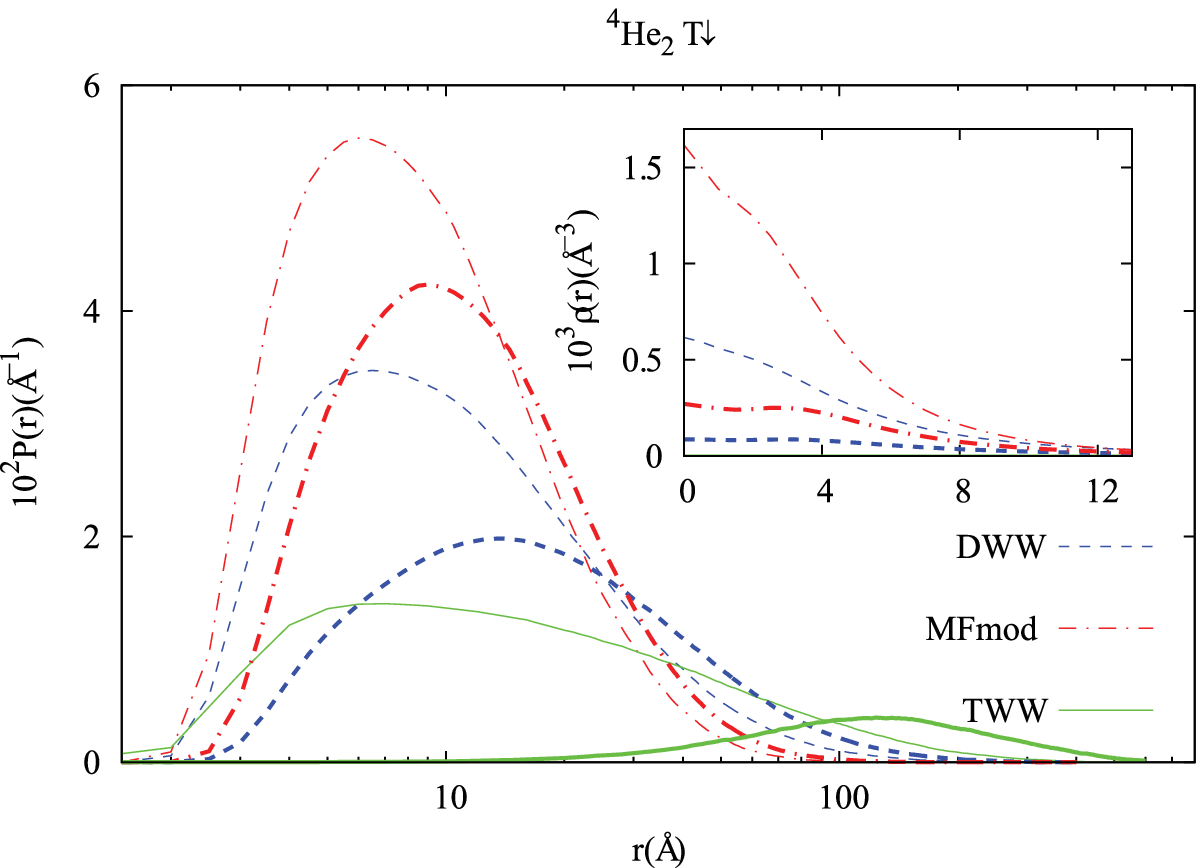}
	\caption{\label{fig:He2T1_pot_comp}Distribution of interparticle distances $P(r)$  for three different interaction potentials, where logscale is used in x-axis. Thinner lines correspond to the distribution of $^4$He-$^4$He distances, and thicker lines to $^4$He-T$\downarrow$ separations. The figure in the inset shows the distributions of the $^4$He (thinner lines) and T$\downarrow$ atoms (thicker lines) to the centre of mass $\rho(r)$ for DWW~\cite{Das} and MFmod~\cite{MFmod} potential. The distribution functions for the TWW~\cite{TWW} potential are not  visible on the picture.}
\end{figure}

In addition to energy, the structure has  also  been determined by calculating
with pure estimators the distribution of particles to the centre of mass of
the system $\rho(r)$ and the distribution of the separations between
particles $P(r)$.  These functions  have been normalized as $\int\rho(r)d^3r=N$, where
$N$ is the number of particles of particular species in the cluster and 
$\int P(r)dr=1$. Both functions, for the smallest cluster
$^4$He$_2$T$\downarrow$ are  shown in  Fig \ref{fig:He2T1_pot_comp}. The
larger plot shows huge differences between distributions of particle
separations for different potentials. As  it  can be expected, for all the
considered potentials two $^4$He atoms are on average closer than one
$^4$He to a T$\downarrow$ atom. The average distance between $^4$He and
T$\downarrow$ atoms is 21.8(5) \AA\, in case of the MFmod potential, 47(2)
\AA\, for the DWW potential and even 206(15) \AA\, for the TWW interaction
potential. At the same time $<r>$ between $^4$He atoms is 74(9)\AA\, for
the TWW potential. This is even larger than the separation of $^4$He atoms
in a $^4$He dimer, which for the selected interaction potential is around
50 \AA. On the other hand, in the case of the MFmod potential  $<r>=
17.0(4)$ \AA\, for the $^4$He-$^4$He pair (30.4(5) \AA\, for the DWW
potential) which confirms the stability of the cluster.  The same floppy
behavior can be noticed from the distribution of particle distances to the
centre of mass, where in the case of the TWW potential the probability of
finding a particle close to the centre of mass is so low  that  it is not visible
in the figure. Correspondingly, the average separation of $^4$He to the
centre of mass is $<r_{cm}>=72(6)$ \AA\, and $<r_{cm}>=150(11)$ \AA\, for
T$\downarrow$, in case of TWW potential, which indicates instability of the
cluster. However, in the course of the simulation there is no indication
that particles are increasing their separation. On the other hand, for 
the 
DWW/MFmod potential $<r_{cm}>$ is 20.1(5) \AA\,/10.2(3) \AA\, for $^4$He
and 32(1) \AA\,/14.3(4) \AA\, for T$\downarrow$. Thus, both energetically
and from the structure analysis $^4$He$_2$T$\downarrow$  is predicted to be
one of the most weakly bound three-atomic clusters.

The differences between the distribution functions for the DWW and MFmod interaction potentials decrease with the increase of the cluster size, but are still noticeable in cases where the number of T$\downarrow$ is about the same or larger than the number of the $^4$He atoms. This can be seen in  Fig. 6 which presents density distributions with respect to the centre of mass  and the $^4$He-$^4$He and $^4$He-T$\downarrow$ separations distribution  for the $^4$He$_2$(T$\downarrow$)$_4$ cluster. 

\begin{figure} 
	\centering
		\includegraphics[width=8.5cm]{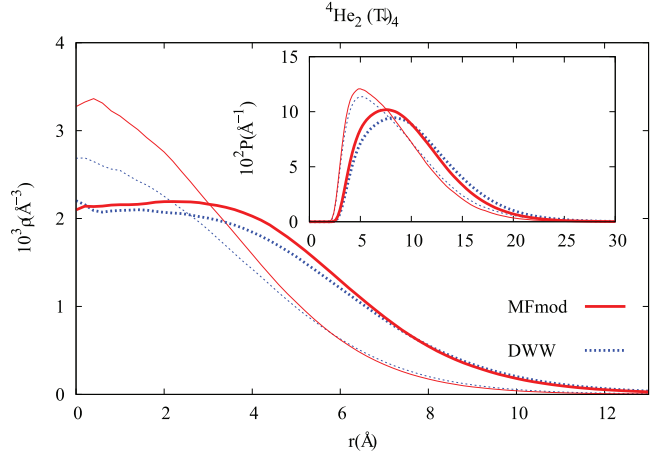}
	\caption{\label{fig:He2T4potcomp}Distribution of of the $^4$He (thinner lines) and T$\downarrow$ atoms (thicker lines) to the centre of mass $\rho(r)$ for DWW~\cite{Das} and MFmod~\cite{MFmod} potentials. The interparticle distance distribution $P(r)$ is shown on the figure in inset. Thinner lines correspond to the distribution of $^4$He-$^4$He distances, and thicker lines to $^4$He-T$\downarrow$ separations. $P(r)$ for T$\downarrow$-T$\downarrow$ are very close to $P(r)$ for $^4$He-T$\downarrow$ and are thus not shown for better clarity. }
\end{figure}
\begin{figure}
	\centering
		\includegraphics[width=8.5cm]{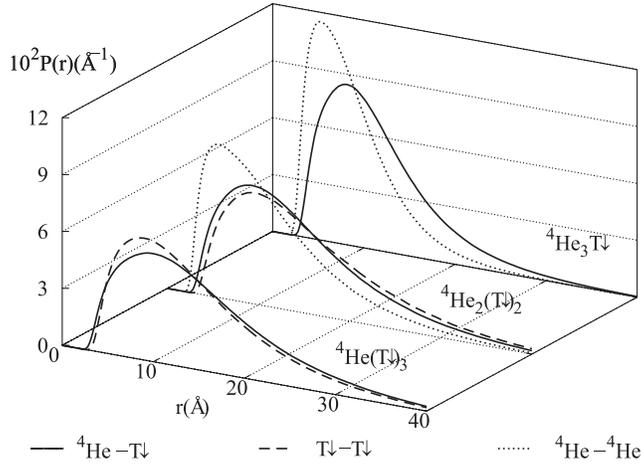}
	\caption{\label{fig:pofr}The distribution of the interparticle distances for three mixed tetramers.  }
\end{figure}

Other small mixed clusters, having in particular 4 to 5 atoms are also very floppy, with typical separations between particles of the order of 10 \AA, for DWW or MFmod potential. Figure \ref{fig:pofr} shows the particle separation distribution for three small mixed four-particle clusters. As expected, two $^4$He atoms have on average the smallest separations. The narrowest distributions correspond to the most strongly bound cluster $^4$He$_3$T$\downarrow$. With the exchange of $^4$He atoms with T$\downarrow$ ones, clusters become more floppy. It is interesting to notice that in the case of $^4$He$_2$(T$\downarrow$)$_2$ and  $^4$He(T$\downarrow$)$_3$ clusters the relationship between particle separations distributions for $^4$He-T$\downarrow$ and T$\downarrow$-T$\downarrow$ pairs is the opposite. 

\begin{figure}
	\centering
		\includegraphics[width=8.5cm]{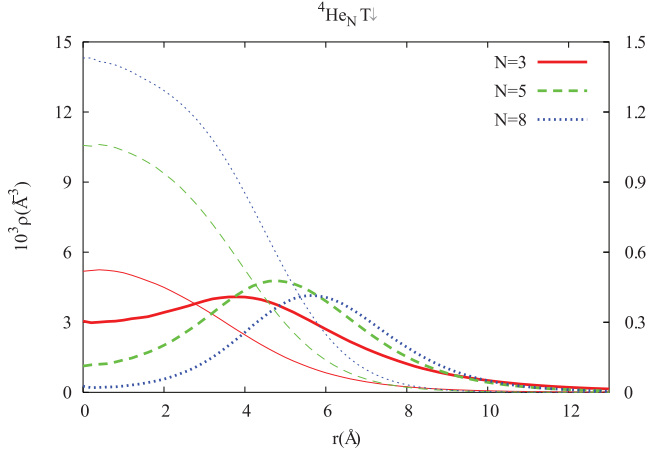}
	\caption{\label{fig:HeTdiscm}Density profiles of the  $^4$He$_N$T$\downarrow$ system, for N=3,5,8. Thin and thick lines stand for $^4$He and T$\downarrow$ atoms, respectively. Left scale is for $^4$He and the right one for T$\downarrow$ density distributions.}
\end{figure}

When T$\downarrow$ is added to the helium cluster $^4$He$_N$, as a consequence of its smaller mass it tends to stay further away from the centre of mass of the system than $^4$He atoms, as can be seen in Fig. \ref{fig:HeTdiscm}. This is completely analogous to the behavior of $^3$He in $^4$He$_N$$^3$He system. The only exception being that T$\downarrow$ is on average further away from the centre of mass in the $^4$He$_N$T$\downarrow$ cluster than $^3$He atom in the $^4$He$_N$$^3$He cluster, which is a consequence of the weaker $^4$He-T$\downarrow$ interaction potential. This can be seen by comparison with the density distribution of $^3$He in Ref. \onlinecite{onehe3}. For $N=3$ there is still appreciable probability for T$\downarrow$  to be close to the centre of mass, while for $N=8$ this probability is already very low, with $<r_{cm}>=7.33(1)$ \AA\ and root-mean-square deviation $\Delta r_{cm}=\sqrt{<r_{cm}^2>-<r_{cm}>^2}$ of 2.2 \AA. For larger clusters, it is expected that T$\downarrow$ will be completely pushed to the surface forming the so-called Andreev states,~\cite{Andreev} like $^3$He in mixed $^3$He-$^4$He clusters.~\cite{nanohelium}

Having determined the  average squared interparticle separations in DMC by
pure estimators, it is interesting to investigate if any of the  smallest 
clusters would fit the definition of the quantum halo state. For three-body
systems a useful definition has been given in Ref. \onlinecite{RMPhalos}.
Quantum halos should obey the condition $\left\langle
\rho^2\right\rangle/\rho_o^2 > 2$, where $\rho$ is the average radial
coordinate and $\rho_0$ a scaling parameter, defined as
\begin{equation}
m\rho^2=\frac{1}{M}\sum_{i<k}m_i m_k(\bm{r_i}-\bm{r_k})^2,~~ m\rho_0^2=\frac{1}{M}\sum_{i<k}m_i m_kR_{ik}^2~.
\label{eq:roovi}
\end{equation}
In the previous equation, $m$ is the arbitrary mass unit, $m_i$ is the mass
of atom $i$, $M$ is the total mass and $R_{ik}$ is the two-body scaling
length of system  $i$ and $k$. For two-body systems $R_{ik}$ can be easily
determined as a classical turning point, from $E=V(r)$. In order to make a
rough qualitative prediction we here assume that on average each pair
contributes with the same amount to the total binding energy. The same
assumption has been made in Ref. \onlinecite{Brunnian}, where clusters of
particles of the same type have been considered. Here, this assumption is
obviously not fully correct, but we have verified that the final
conclusions whether a cluster is a candidate for halo state or not, remain
the same if we assume absolute average potential energies of some pairs to
be larger and others smaller. Finally, since the classical region is
defined by having positive kinetic energy, we estimate the classical radius
by equaling the total energy divided by the number of pairs with the
corresponding potential energy of each pair. Having thus found all the
$R_{ik}$ we calculate the scaling parameter $\rho_0$ from Eq.
(\ref{eq:roovi}). The only bound mixed trimer $^4$He$_2$(T$\downarrow$)
fits the definition of the quantum halo state since $\left\langle
\rho^2\right\rangle/\rho_o^2 = 13.6$. The result is insensitive to
different types of $^4$He-T$\downarrow$ interaction potentials. The authors
of Ref. \onlinecite{Brunnian} have used the same approach to estimate the
halo condition for clusters of four and five particles. If we extend  Eq.
(\ref{eq:roovi}) to larger clusters we find that both
$^4$He$_2$(T$\downarrow$)$_2$ and  $^4$He(T$\downarrow$)$_3$ can be
considered quantum halo states in case of the DWW $^4$He-T$\downarrow$
interaction potential, since  $\left\langle
\rho^2\right\rangle/\rho_o^2$  is  2.32 and
2.04, respectively. This approach is
qualitative, but it confirms the extremely weak
structure of these
systems and motivates their further investigation.

\section{conclusions}
The ground state of small mixed clusters composed of helium and
spin-polarized tritium has been investigated by quantum Monte Carlo
simulations. Significant differences in binding energies are obtained for different $^4$He-T$\downarrow$
interaction potentials,  especially for the smallest clusters. Nevertheless, the conclusions concerning stability
limits are insensitive to these differences, except for the trimer. The
mixed dimer does not exist, and the only bound mixed trimer is the
$^4$He$_2$T$\downarrow$ one. The latter  result
  is obtained for several
$^4$He-T$\downarrow$ interaction potentials. However, in   the  case of the
weakest potential (TWW) the cluster  $^4$He$_2$T$\downarrow$ is at 
threshold of separating into a dimer and a free particle and its stability
could not be determined with certainty. Due to its large size, this cluster
has most of the probability outside the classically allowed regions of
space, classifying it as a quantum halo state. Another possible mixed
trimer $^4$He(T$\downarrow$)$_2$, a candidate for the Borromean state is
not bound. Our results further indicate that two mixed tetramers
$^4$He$_2$(T$\downarrow$)$_2$ and  $^4$He(T$\downarrow$)$_3$, could also be
considered quantum halos. The largest clusters here analyzed are also very
weakly bound and floppy. 

\acknowledgments

We thank I. Be\v{s}li\'c for useful discussions.
J. B. acknowledges support from DGI (Spain) Grant No.
FIS2005-04181 and Generalitat de Catalunya Grant No. 2008SGR-04403. P.S. and L.V.M. acknowledge
support from MSES (Croatia) under Grant No. 177-1770508-0493. 
We also acknowledge the support of the Central Computing Services at
the Johannes Kepler University in Linz, where part of the computations was
performed.  In addition, the resources of the Isabella
cluster at Zagreb University Computing Centre (Srce) and Croatian National
Grid Infrastructure (CRO NGI) were used, as well as the resources of the HYBRID cluster at the University of Split, Faculty of Science.

\end{document}